# Magnetic-dipolar-mode Fano resonances for microwave spectroscopy of high absorption matter

G. Vaisman, E.O. Kamenetskii, and R. Shavit

Microwave Magnetic Laboratory,
Department of Electrical and Computer Engineering,
Ben Gurion University of the Negev, Beer Sheva, Israel

January 16, 2015

**Abstract**

Study of interaction between high absorption matter and microwave radiated energy is a subject of great importance. Especially, this concerns microwave spectroscopic characterization of biological liquids. Use of effective testing methods to obtain information about physical properties of different liquids on the molecular level is one of the most important problems in biophysics. However, the standard methods based on the microwave resonant techniques are not sufficiently suitable for biological liquids because the resonance peak in a resonator with high-loss liquids is so broad that the material parameters cannot be measured correctly. Although molecular vibrations of biomolecules may have microwave frequencies, it is not thought that such resonant coupling is significant due to their low energy compared with thermal energy and the strongly dampening aqueous environment. This paper presents an innovative microwave sensing technique for different types of lossy materials, including biological liquids. The technique is based on the combination of the microwave perturbation method and the Fano-resonance effects observed recently in microwave structures with embedded magnetic-dipolar quantum dots. When frequency of the magnetic-dipolar-mode (MDM) resonance is not equal to the cavity resonance frequency, one gets Fano transmission intensity. With tuning, by a bias magnetic field, the MDM resonance frequency to the cavity resonance frequency, one observes a Lorentz line shape. Use of an extremely narrow Lorentzian peak allows exact probing the resonant frequency of a cavity loaded by a high-lossy-material sample. For different kinds of samples, one has different frequencies of Lorentzian peaks. This gives a picture for precise spectroscopic characterization of high absorption matter in microwaves.

**I. INTRODUCTION**

Fano resonances arise as two transmission pathways, a broad band continuum and a narrow band resonance, interfere with each other. Being originated in atomic physics [1], Fano resonances have become one of the most appealing phenomena in semiconductor quantum dots [2 – 4], different photonic devices [5 – 10], and microwave structures [11 – 13]. An interest in observing and analyzing Fano profiles is driven by their high sensitivity to the details of the scattering process. The Fano-resonant phenomenon appears as a notch in the absorption spectrum when the incident electromagnetic wave couples to a strongly damped oscillator, which in turn is coupled to a weakly damped mode. The resulting effective coupling between the two modes is dispersive, i.e., it depends strongly on the frequency in a narrow interval around the frequency of the weakly damped oscillator and gives rise to a strong modulation of the absorption spectrum. Presently, these effects are widely used in optical spectroscopy of biological structures [8, 9]. At the same time, there are no publications, to the best of our knowledge, on the Fano-resonant spectroscopy of biological structures in microwaves.

Study of microwave properties of biological liquids is one of the most important problems in biophysics. Microwave absorption is primarily a tool for observing and measuring different kinetic processes in biological liquids. It may concern molecule rotational transitions and vibrational resonances in biological systems. However, such microwave excitable resonances are expected to be strongly damped by interaction with their aqueous biological environment. In literature, there are numerous discussions regarding microwave absorption resonances in aqueous solutions of DNA. In Ref. [14], Edwards *et al* stated that they experimentally observed the resonant absorption of microwave energy by aqueous solutions containing DNA. Based on a simple model of dissolved DNA, van Zandt [15] gave a theoretical explanation of anomalous microwave absorption resonances observed in Ref. [14]. These results, however, were considered as controversial. In Refs. [16, 17], it was demonstrated that there are no microwave resonance effects owing to DNA. An extended analysis in Ref. [18] also suggests that the damping of the vibration motion by biological fluids severely restricts possibilities to observe such resonances.

The success of a spectroscopic technique depends on two main problems: one is the availability of accurate data over the required frequency range while the other is the ability to unambiguously interpret this data. These two problems have a special aspect in spectroscopic characterization of high lossy materials. The temporal and spatial resonant nature of an electromagnetic standing wave within a dissipative media is obscure and incoherent. Spatially, regions with maximum magnetic (electric) field and null electric (magnetic) field can no longer be well defined and separated. One of the topical subject on the lossy-material characterization concerns microwave analyses of biological liquids. Proper correlation of the measured parameters with structural characteristics of chemical and biological objects in microwaves appears as a serious problem. Nowadays, microwave biosensing is mainly represented by the microwave-cavity technique and the transmission/reflection technique [19 – 26]. The microwave-cavity technique is based on the well known perturbation method used for measuring dielectric properties of materials. The resonant techniques, however, are not suitable for high-lossy liquids because the resonance peak is so broad that the perturbation characteristics cannot be measured correctly. Also, interpretation the data obtained from the transmission/reflection technique cannot be considered as sufficiently suitable for high-lossy liquids.

In numerous microwave experiments, a problem of the high-lossy-material characterization is formulated so that to obtain data on the permittivity and permeability parameters which are represented as a combination of real and imaginary parts – $\varepsilon'$, $\varepsilon''$ and $\mu'$, $\mu''$. It appears, however, that such representation of parameters and, moreover, interpretation that these data properly characterize high-lossy materials are beyond a physical meaning. It is known that electromagnetic processes in a non-conductive medium are described, energetically, by Poyning's theorem which, being represented in a form of the continuous equation:

$$\vec{\nabla} \cdot \vec{S} = \frac{\partial w}{\partial t} + p, \qquad (1)$$

shows that divergence of the power flow density $\vec{S}$ is determined by two quantities: time variation of the electromagnetic-field density $w$ and the density of the dissipation losses $p$. For an isotropic temporally dispersive medium, described by constitutive parameters $\varepsilon(\omega) = \varepsilon'(\omega) + i\varepsilon''(\omega)$ and $\mu(\omega) = \mu'(\omega) + i\mu''(\omega)$, the two terms in the right-hand side of Eq. (1) have definite physical meaning (and so can be considered separately) only when

$$|\varepsilon''(\omega)| << |\varepsilon'(\omega)| \text{ and } |\mu''(\omega)| << |\mu'(\omega)|. \tag{2}$$

The frequency regions where $|\varepsilon''(\omega)|$ and $|\mu''(\omega)|$ are small compared to $|\varepsilon'(\omega)|$ are $|\mu'(\omega)|$ are called the regions of transparency of a medium. Only inside transparency regions one can separately introduce a notion of the electromagnetic-field density $w$ (which is related to the real-part quantities $\varepsilon'$ and $\mu'$) and a notion of the density of the dissipation losses $p$ (which is related to the imaginary-part quantities $\varepsilon''$ and $\mu''$). The energy balance equation (1) for such a transparent medium is described by quasi-monochromatic fields [27, 28]. For a non-transparent medium (or in a frequency region of medium non-transparency), relations (2) are unrealizable and one cannot describe the right-hand side of Eq. (1) by two separate and physically justified terms. This is the case of a high lossy (or high absorption) medium. In a high lossy medium, one cannot state that the electromagnetic-field density $w$ and the density of the dissipation losses $p$ are separate notions with definite physical meaning. In such a case, there is no meaning to consider quantities $\varepsilon'$ and $\mu'$ as physical parameters related to energy accumulation and quantities $\varepsilon''$ and $\mu''$ as physical parameters related to energy dissipation.

In all the experiments for characterization of high-lossy material parameters, one does not use the quasi-monochromatic fields. Decay and a frequency shift, observed in these microwave experiments, are the quantities which describe the high-lossy materials very indirectly. Two basic strategic questions are thus posed: (a) What is the physical meaning of the constitutive parameters characterizing electromagnetic processes in high lossy materials? and (b) How these parameters can be precisely measured? The fact that in known standard microwave experiments, high-lossy material characteristics not only can be measured accurately, but also cannot be interpreted physically corrected (assuming that there are the real and imaginary parts of the permittivity and permeability parameters), arise a question on possibilities of realization of novel and appropriate techniques. In this paper we propose such a technique. This technique is based on the Fano-interference effect in a microwave structure with an embedded small ferrite-disk resonator.

The use of a small ferrite-disk scatterer with internal magneto-dipolar-mode (MDM) resonances in the channel of microwave propagation changes the transmission dramatically. Recently, it was shown that mesoscopic quasi-2D ferrite disks, distinguishing by multiresonance MDM oscillations, demonstrate unique properties of artificial atomic structures: energy eigenstates, eigen power-flow vortices and eigen helicity parameters [29 – 35]. These oscillations can be observed as the frequency-domain spectrum at a constant bias magnetic field or as the magnetic-field-domain spectrum at a constant frequency. For electromagnetic waves irradiating a quasi-2D MDM disk, this small ferrite sample appears as a topological defect with time symmetry breaking. The physical meaning of such a topological singularity is the following. Recent studies in optics show that for a case of a nanoparticle illuminated by the electromagnetic field, the 'energy sink' vortices with spiral energy flow line trajectories are seen in the proximity of the nanoparticle's plasmon resonance [36]. In microwaves, ferrite particles with the ferromagnetic-resonance (FMR) conditions provoke creation of electromagnetic vortices as well. In papers [37, 38], ferrite inclusions with the FMR conditions were used to analyse a role of the time-reversal symmetry breaking as a factor leading to creation of the Poynting-vector microwave vortices. When such ferrite inclusions are thin-film ferrite disks, one can observe eigen power-flow vortices of MDM oscillations [31 – 35]. Long radiative lifetimes of topologically singular MDMs combine strong subwavelength confinement of electromagnetic energy with a narrow spectral line width and may carry the signature of Fano resonances [39 – 41]. It was shown [42] that interaction of the MDM ferrite particle with its environment has a deep analogy with the Fano-resonance interference observed in natural and artificial atomic structures. The aim of the present study is to show how the transition between symmetric Lorentzian and asymmetric Fano line shapes in microwave structures with MDM ferrite disks can be used for effective microwave spectroscopy.

In this paper, we use a rectangular-waveguide microwave cavity with low-quality factor $Q$ as a strongly damped oscillator, while a MDM ferrite disk (embedded into a microwave cavity) manifests itself as a weakly damped oscillator. Being a wave interference phenomenon, Fano resonance is highly sensitive to the scattering details of a system and can be used in the study of various transport properties. In our studies, variation of the transport properties is due to different lossy samples loading the microwave cavity. Effective coupling between the two modes (strongly damped and weakly damped ones) results in a strong modulation of the cavity field distribution and the cavity absorption spectrum. The lineshape of the weakly damped peaks highly depends on the frequency of the MDM oscillation. Also, the field topology in the cavity is different for different MDM frequencies. The resonant frequency of MDM oscillations depends on the magnitude of a bias magnetic field. When frequency of the MDM resonance is not equal to the cavity resonance frequency, one gets an asymmetric Fano lineshape of the peaks. With tuning, by a bias magnetic field, the MDM resonance frequency to the cavity resonance frequency, one can obtain a symmetric Lorentz lineshape of the peaks. Use of an extremely narrow Lorentzian peak allows exact probing the resonant frequency of a cavity loaded by a high-lossy-material sample. The cavity resonant frequency is the frequency of minimum power absorption of the cavity. For different kinds of samples, one has different frequencies of Lorentzian peaks. This gives a picture for precise spectroscopic characterization of high absorption matter in microwaves.

## II. A MDM FERRITE DISK AND A MICROWAVE CAVITY: THE MODEL OF COUPLED LORENTZIAN OSCILLATORS

MDM oscillations in a quasi-2D ferrite disk are macroscopically quantized states. Long range dipole-dipole correlation in position of electron spins in a ferromagnetic sample can be treated in terms of collective excitations of the system as a whole. If the sample is sufficiently small so that the dephasing length $L_{ph}$ of the magnetic dipole-dipole interaction exceeds the sample size, this interaction is non-local on the scale of $L_{ph}$. This is a feature of mesoscopic ferrite samples, i.e., samples with linear dimensions smaller than $L_{ph}$ but still much larger than the exchange-interaction scales. MDMs in a quasi-2D ferrite disk possess unique physical properties. Being the energy-eigenstate oscillations, they also are characterized by topologically distinct structures of the fields. There are the rotating field configurations with power-flow vortices. At the MDM resonances, one observes power-flow whirlpools in the vicinity of a ferrite disk. For an incident EM wave, such a vortex topological singularity acts as a trap, providing strong subwavelength confinement and symmetry breakings of the microwave field [29 – 35, 43, 44].

For a MDM ferrite particle placed inside a microwave cavity with sufficiently low quality factor $Q$, one can observe the Fano-interference effects in microwave scattering. The spectrum of the MDM oscillations in a ferrite-disk particle is very rich [45, 46]. It contains different types of modes: the radial, azimuthal, and thickness modes [30]. In the present study, we consider only the case of interaction of the cavity oscillation with the main mode in the MDM-spectrum sequence. Fig. 1 (*a*) shows a rectangular-waveguide microwave cavity with an enclosed yttrium-iron-garnet (YIG) disk. A ferrite disk has a diameter of $2\Re = 3 \text{ mm}$. The YIG film thickness is $d = 49.6 \text{ um}$. Saturation magnetization of a ferrite is $4\pi M_0 = 1880 \text{ G}$ and the linewidth is $\Delta H = 0.8 \text{ Oe}$. The sizes of our ferrite sample essentially exceed the exchange-interaction scales, which are about units of micrometers and less. A normally magnetized ferrite-disk sample is placed in a rectangular waveguide cavity with the $TE_{102}$ resonant mode. This resonant mode has two antinodes of the electric-field intensity along z axis. The disk is placed between these electric-field antinodes, where one has an antinode of the cavity magnetic field. The disk axis is oriented along the waveguide *E*-field and the disk position is in a maximum of the RF magnetic field of the cavity. Fig. 1 (*b*) shows the numerically obtained (by the HFSS program) frequency characteristics of a transmission coefficient in the structure. With tuning, by a bias magnetic field, the MDM resonance frequency one observes strong transformation of lineshape of the peaks. At frequencies of the MDM resonance not equal to the cavity resonance frequency, one gets Fano transmission intensity. When frequency of the MDM resonance is exactly equal to the cavity resonance frequency, one observes a Lorentz line shape.

From Fig. 1 (*b*), one can see that coupling between a MDM resonance and electromagnetic (EM) cavity resonance results in appearance of destructive and constructive interferences. At the destructive interferences, the channels of EM waves propagating in a microwave cavity are effectively suppressed and a transmission coefficient in the structure is sharply reduced. There is a purely interference effect which is not related to the absorption by MDM oscillations. Since the uncoupled MDM resonance is extremely sharp, as compared to the EM-cavity-resonance width, we can set the MDM damping equal to zero. The observed quantum-like interference between the excitation pathways, that control the EM response, can be classified as electromagnetically induced non-transparency (in contrast to a well known effect of electromagnetically induced transparency [47]). In a case of the constructive interferences, a level of a transmission coefficient in the structure is slightly increased.

The destructive and constructive interference due to coupling between MDM and EM-cavity resonances manifests itself by the field topology. The topological structures of the fields in a cavity, shown in Fig. 2, give evidence for the fact that in a case of constructive interference the fields almost correspond to a regular $TE_{102}$ resonant mode and a ferrite disk slightly perturbs the power-flow-density distribution in a cavity. Contrarily, for destructive interference, a ferrite disk strongly localizes energy. One observes a characteristic picture of the Poynting-vector vortices near a ferrite disk. The field topology is strongly disturbed, compared to the fields of a regular $TE_{102}$ resonant mode. Fig. 3 shows dynamics of the field structure for a case of destructive interference.

A system of coupled MDM and EM-cavity resonators can be well described by a classical model of coupled Lorentzian oscillators. Such a non-conservative system can be represented as two coupled *LC* circuits or two mechanical oscillators consisting on masses and strings [48 – 52]. Following along the lines of Ref. [49], we write the equations of motion for the system of two coupled oscillators. For two classical oscillators, *A* and *B*, excited by a source $F(t)$, the equations are:

$$\ddot{x}_A + \gamma_A \dot{x}_A + \omega_A^2 x_A + c x_B = g_A F(t) , \qquad (3)$$

$$\ddot{x}_B + \gamma_B \dot{x}_B + \omega_B^2 x_B + c x_A = g_B F(t), \qquad (4)$$

where $x_j(t)$ ($j = A, B$) are the displacements representing the oscillations; $\omega_j$, $\gamma_j$, and $g_j$ ($j = A, B$) are, respectively, the resonance frequencies, damping rates, and oscillation strengths of the uncoupled oscillators; $c$ represents the coupling strength between the oscillators. Let oscillator $A$ be a highly damped EM-cavity resonator and oscillator $B$ a narrow band MDM resonator. Since the uncoupled MDM resonance is extremely sharp as compared to the EM-resonance width, we can assume that $\gamma_B \ll \gamma_A$. For the same reason, we can also assume that $|g|_B \ll |g|_A$. In a further analysis, we set the MDM damping $\gamma_B = 0$. Regarding the oscillation strength coefficients $g_j$, most aspects can already be understood in the simpler case $g_B = 0$. This case – the case when coupled oscillators are driven by a force applied to one of the oscillators – well describes interaction of the cavity oscillation with the main mode in the MDM-spectrum sequence. At the same time, for an analysis of coupling of high-order MDMs with the EM cavity, small but finite quantities of the coefficient $g_B$ should be taken into account when strong overlapping between high-order MDMs occurs. This problem of high-order MDMs is beyond the frames of the present study.

For the frequency domain, Eqs. (3) and (4) can be solved analytically. In monochromatic excitation, i. e. for $F(t) = f e^{i\omega t} + \text{c.c}$, harmonic responses of the displacements are represented as $x_j(t) = X_j e^{i\omega t} + \text{c.c}$. Assuming that $g_B = 0$, we are interested only in the power absorbed by oscillator $A$ from a source. So, we will seek a solution for $x_A(t)$. From Eqs. (3), (4), and with a condition that $\gamma_B = 0$, we obtain for complex amplitude $X_A$:

$$X_A = \frac{\omega_B^2 - \omega^2}{(\omega_A^2 - \omega^2 + i\gamma_A \omega)(\omega_B^2 - \omega^2) - c^2} f . \tag{5}$$

In our analysis we are looking for solutions in a frequency interval near the frequency of the oscillator $B$. Around the frequency $\omega_B$, the two oscillators interfere destructively and constructively, giving rise to an asymmetrical resonance lineshape. Now, following a general-form analysis in Ref. [52], we make the Fourier transform of Eqs. (3) and (4). This transform gives:

$$\mathcal{L}_\omega X_A + c X_B = \tilde{F}, \tag{6}$$

$$\left(\omega_B^2 - \omega^2\right) X_B + c X_A = 0, \tag{7}$$

where operator $\mathcal{L}_\omega$ is defined as $\mathcal{L}_\omega \equiv \omega_A^2 + i\gamma_A \omega - \omega^2$; $X_j$ and $\tilde{F}$ are the Fourier components of the displacements $x_j(t)$ and the source $F(t)$. For simplicity, we take here $g_A = 1$. In an assumption that $\mathcal{L}_\omega$ is slowly varying in a frequency interval around $\omega_B$, we obtain for the complex amplitude $X_A$

$$X_A \approx \frac{\omega_B^2 - \omega^2}{\mathcal{L}_{\omega_B}(\omega_B^2 - \omega^2) - c^2} \tilde{F}, \tag{8}$$

where $\mathcal{L}_{\omega_B} \equiv (\mathcal{L}_\omega)|_{\omega=\omega_B}$. With some algebraic manipulations, shown in Ref [38], we obtain from Eq. (8):

$$|X_A|^2 = \frac{|\tilde{F}|^2}{|\mathcal{L}_{\omega_B}|^2} \frac{(\kappa+q)^2}{\kappa^2+1}. \tag{9}$$

The parameter $q$, describing the degree of asymmetry of the line shape, is defined as $q = -\text{Re}(\mathcal{L}_{\omega_B})/\text{Im}(\mathcal{L}_{\omega_B})$. The reduced-frequency parameter $\kappa$ is defined by $\kappa = (\omega^2 - \omega_b^2 - \omega_b \Delta)/\Gamma$, where $\Delta = -\frac{c^2}{|\mathcal{L}_{\omega_B}|^2}\frac{\text{Re}(\mathcal{L}_{\omega_B})}{\omega_B}$ and $\Gamma = \frac{c^2}{|\mathcal{L}_{\omega_B}|^2}\text{Im}(\mathcal{L}_{\omega_B})$. The term $\frac{(\kappa+q)^2}{\kappa^2+1}$ in Eq. (9) is the classical equivalent of the Fano formula [39] around the resonance frequency of the oscillator $B$. As the oscillator $B$ is subjected to a frequency shift of the phase difference $\pi$ around its resonance frequency, one has destructive or constructive interference. For $\kappa = -q$, the destructive interference leads to complete vanishing of $|X_A|^2$.

In the Fano formula, the parameter $q$ describes the degree of asymmetry of the resonance. For $q > 0$, the amplitude of a broad band channel exhibits a sharp transition from minimum to maximum. The opposite situation takes place for $q < 0$. When $q = 0$, the Lorentzian shape occurs. For pure harmonic oscillations, a sign of parameter $q$ is determined by a sign of difference $\omega_B^2 - \omega_A^2$ [52]. For $\omega_B > \omega_A$, we have $q > 0$, while for $\omega_B < \omega_A$, there is $q < 0$. In a case of $\omega_B = \omega_A$, parameter $q$ is equal to zero [it is worth noting here that parameter $\kappa$ in Eq. (9) also is equal to zero when $\omega_B = \omega_A$]. The above classical model of coupled Lorentzian oscillators well describes frequency characteristics of the transmission coefficient of a system of coupled MDM and EM-cavity resonators.

## III. MICROWAVE FANO-RESONANT SPECTROSCOPY FOR HIGH-ABSORPTION-MATTER CHARACTERIZATION: MATERIALS AND METHODS

In widely used microwave cavity technique to determine the complex permittivity material parameters, the perturbation approach is commonly applied, which is characterized with limitation on permittivity and losses values as well as sample dimensions. The perturbation theory requires that sample permittivity, losses values and dimensions should be small enough so that the field distribution inside the empty cavity changes slightly when the cavity is loaded. Only with this limitation, the perturbation technique permits to link via a simple formula changes in the resonant frequency and loaded factor determined by the sample.

As an example of such a formula, there is [53]:

$$\frac{\Delta f}{f_0} + \frac{1}{2}i\Delta\frac{1}{Q} = k\frac{-\chi_e}{(1+F_{sh}\chi_e)}, \tag{10}$$

where $\chi_e = \chi_e' - i\chi_e''$ is the complex dielectric susceptibility, $f_0$ is the frequency of an unloaded cavity, $\Delta f$ and $\Delta \frac{1}{Q}$ are, respectively, the frequency and quality-factor shifts due to dielectric loading, $F_{sh}$ is the constant dependent on the sample shape, and $k$ is a calibration constant. One can see that both the frequency shift and the quality-factor shift depend on both the real and imaginary parts of the dielectric susceptibility. Separate experimental evaluation of these two kinds of shifts is possible only for small material losses. For high-lossy materials, the cavity resonance curve becomes so wide that evaluation of the quality-factor shift cannot presume exact evaluation of the frequency shift. Moreover, since the cavity resonance curve is wide, other resonances (originated, for example, from the terminal connectors) become prominent and an accurate detection of the resonant frequency shift appears to be very difficult.

Based on the Fano-interference effect in a microwave structure with an embedded small ferrite-disk resonator, we present an innovative microwave-cavity technique suitable for precise spectroscopical characterization of different types of lossy materials, including biological liquids. An experimental structure is shown in Fig. 4. Fig. 4 (*a*) shows the experimental setup for characterization of high-absorption-matter parameters based on MDM Fano resonances. In 4 (*b*), one can see a device-under-test (DUT): a microwave cavity with embedded a small ferrite disk and a material sample. A bias magnetic field $\vec{H}_0$ is produced by using an electromagnet. A power supply (LAMBDA dc power supply) feeds the electromagnet with a DC current that can be adjusted (using LabView) to produce the desired bias magnetic field. The bias field is measured with a gauss meter (2010T, Magnetic Instrumentation Co.) with accuracy of 0.05 Oe. As the RF source we used the vector network analyzer (Agilent 87050, power level of -15dBm, Agilent Co.). Spurious signals are negligible since there are no active devices in our setup. Accuracy in our microwave measurements is 150 KHz. Our experimental studies are made with use of a ferrite structure similar to the structure for numerical studies. In experiments, we use a disk sample (Ferrite Domen Co.) of diameter $2\Re = 3\,\text{mm}$ made of the yttrium iron garnet (YIG) film on the gadolinium gallium garnet (GGG) substrate (the YIG film thickness $d = 49.6\,\mu\text{m}$, saturation magnetization $4\pi M_0 = 1880\,\text{G}$, linewidth $\Delta H = 0.8\,\text{Oe}$; the GGG substrate thickness is 0.5 mm). A normally magnetized ferrite-disk sample is placed in a rectangular-waveguide cavity with the TE$_{102}$ resonant mode. The disk axis is oriented along the waveguide *E*-field and the disk position is in a maximum of the RF magnetic field of the cavity. A cavity with a ferrite disk is loaded by a dielectric sample placed in a maximum of the RF electric field of the cavity. A transmission coefficient (the $S_{21}$ parameter of the scattering matrix) was measured with use of a network analyzer. With use of a current supply we established and tuned a bias magnetic field $\vec{H}_0$. A quantity of a bias magnetic field was measured by a gaussmeter. In the experiments, the ferrite disk and the samples held in place with use of styrofoam holders. Relative permittivity of a styrofoam material is about unit and microwave losses are very small so such holders do not practically influence on our measurements.

We start our experiments with demonstration of the above mentioned problems of the material characterization by a standard perturbation technique. Without a bias magnetic field $\vec{H}_0$, a ferrite disk behaves as a small dielectric sample. Since a ferrite disk is placed in a maximum of the cavity RF magnetic field, its role (when $\vec{H}_0 = 0$) is negligibly small. Fig. 5 (*a*) shows the frequency characteristics of a transmission coefficient ($S_{21}$ parameter of the scattering matrix) when a cavity is loaded with small low-loss dielectric samples. There are two ceramics disks (diameter of 3 mm and thickness of 2 mm) with the dielectric constants $\varepsilon_r = 30$ (K-30; TCI Cermics, Inc.) and $\varepsilon_r = 50$ (K-30; TCI Cermics, Inc.). For such dielectric samples, the cavity mode is highly coherent and other "parasitic" scattering channels are less dominant, making the standard perturbation technique applicable. In this case, the cavity's quality factor is sufficiently big to detect the resonance shift and there is no need to use a ferrite disk for probing the exact resonance frequency. Furthermore, even if a bias magnetic field is not zero, the Fano-resonance effect does not occur in this case since the cavity field cannot be considered to vary slowly at the vicinity of the MDM resonance band. In Fig. 5 (*b*), we show the frequency characteristics of a transmission coefficient when the same dielectric samples ($\varepsilon_r = 30$ and $\varepsilon_r = 50$) are wrapped by a polyurethane. A polyurethane material is a polymer composite material widely used as an effective absorber of microwave radiation. We used a single layer structure of this material. In a case of a dielectric sample with a polyurethane structure, a cavity quality factor is strongly reduced. The $TE_{102}$ mode becomes less dominant and resonance frequency shifts are not as clear as in the low loss scenario. Certainly, an exact characterization of the sample parameters by transformation of the spectral characteristics is impossible in this case.

Now we illustrate how the proposed Fano-interference technique can be used to detect the cavity resonance frequency when the cavity is loaded by a high-lossy material. The experimental characteristic shown in Fig. 6 is obtained when a cavity is loaded only by a high-lossy wrapping material. When a proper bias magnetic field is switched on, the MDM spectral peaks of a ferrite disk are observed. Fig. 6 shows the peak positions of the main MDM for different quantities of a bias magnetic field. It is observed clearly that while for magnetic fields $\vec{H}_0^{(1)}$ and $\vec{H}_0^{(2)}$ we have the same type of the peak lineshape corresponding to the Fano asymmetry parameter $q > 0$, for a bias field $\vec{H}_0^{(4)}$ there is another type of the peak lineshape with the parameter $q < 0$. A bias magnetic field $\vec{H}_0^{(3)}$ gives us exactly a Lorentzian-like shape ($q = 0$). Change of a sign of the Fano asymmetry parameter is due to coupling between a narrow band resonator (a MDM ferrite disk) and a highly damped resonator (the $TE_{102}$-mode cavity with a lossy-material sample). A ferrite disk is not coupled with other scattering channels. For $\vec{H}_0^{(3)} \equiv \vec{H}_0^{(Lorentz)}$ the resonance frequencies of the MDM disk and $TE_{102}$-mode cavity are equal. So, an external-parameter quantity $\vec{H}_0^{(3)} \equiv \vec{H}_0^{(Lorentz)}$ marks exactly the cavity resonance frequency.

The proposed Fano-interference technique can be used for precise spectroscopical characterization of high-lossy materials. This technique does not give us exact values of parameters $\varepsilon'$, $\varepsilon''$ and $\mu'$, $\mu''$. As we discussed above, the physical meaning of these parameters is not clear for high-lossy-material samples. In our studies, we can mark exactly the position of a very narrow Lorentz peak on the frequency axis corresponding to a certain type of a material. For different types of materials, one has different spectroscopical Lorentz-peak markers on the frequency axis. As different types of high-lossy-material samples, we use the complex structures of dielectric samples ($\varepsilon_r = 30$ and $\varepsilon_r = 50$) wrapped by a polyurethane material. The samples are placed in a maximum of the RF electric field of the cavity. For $\vec{H}_0 = 0$ the frequency characteristics of a transmission coefficient in a cavity are shown in Fig. 5 (*b*). When we switch on a bias magnetic magnetic field, these characteristics remains generally the same, but sharp MDM peaks appear additionally. With variation of a bias field one obtains the necessary Lorentz-peak bias field for every type of a high-lossy-material sample. The frequency of such a Lorentz-peak marks exactly the cavity resonance frequency for a certain material loading this cavity. Fig. 7 (*a*) gives spectroscopical pictures for high-lossy-material samples. The pictures are obtained for the main MDM of a ferrite disk. To increase sensitivity of our spectroscopical analysis, we normalize the frequency characteristics to the background (when $\vec{H}_0 = 0$) transmission characteristics. Fig. 7 (*b*) shows the normalized spectroscopical pictures.

The possibility to use the proposed Fano-interference technique for spectroscopical signature of liquids is well illustrated by our experiments with salt solutions and glucose solutions. These solutions represent a standard basis of biological samples. Fig. 8 (*a*) shows the frequency characteristics of a transmission coefficient of a cavity loaded by a small capsule with distilled water and the NaCl salt solutions. The capsule is a small cylinder (length of 3mm, internal diameter of 1.5mm, and wall thickness of 0.2mm) made of polyethylene. The capsule with solutions is placed in a maximum of the RF electric field of the cavity and a bias magnetic field $\vec{H}_0 = 0$. One can see that there are no possibilities for precise spectroscopical characterization of these solutions. The Lorentz-peak marks shown in Figs. 8 (*b*) and (c) allow exact characterization of the salt concentration. These marks are the cavity resonance frequencies obtained for certain values of a bias magnetic field. In Figs. 9 (a), (*b*) and (c) one can see a similar procedure for spectroscopical signature of glucose solutions.

## IV. CONCLUSION

The standard methods based on the resonant techniques cannot characterize precisely microwave properties of high absorption matter. Especially, this concerns microwave spectroscopic characterization of biological liquids. In a microwave resonator loaded by high-lossy liquid, the resonance peak is so broad that the material parameters cannot be measured correctly. In this paper we propose a novel microwave technique for precise spectroscopical characterization of high-lossy materials. This technique is based on the Fano-interference effect in a microwave structure with an embedded small ferrite-disk resonator.

The use of a small ferrite-disk scatterer with internal MDM resonances in the channel of microwave propagation changes the transmission dramatically. In this paper we showed that based on a combination of the microwave-cavity perturbation method and the Fano-resonance interference in microwave structures with embedded magnetic-dipolar dots, one gets a unique microwave sensing technique for precise spectroscopical characterization of different-type lossy materials, including biological liquids. With tuning, by a bias magnetic field, the MDM resonance frequency to the cavity resonance frequency, one observes a symmetric Lorentz lineshape of the peaks. Importantly, we distinguish the cavity resonant frequency as the frequency of a minimum of power absorption in the cavity. So, for precise spectroscopic characterization of high-lossy materials there is no need for separate estimations of complex permittivity and permeability parameters used in traditional techniques.

Use of an extremely narrow Lorentzian peak allows exact probing the resonant frequency of a cavity loaded by a high-lossy-material sample. For different kinds of samples, one has different frequencies of Lorentzian peaks. With variation of a bias field one obtains the necessary Lorentz-peak bias field for every type of a high-lossy-material sample. The frequency of such a Lorentz-peak marks exactly the cavity resonance frequency for a certain material loading this cavity. The possibility to use the proposed Fano-interference technique for spectroscopical signature of liquids is well illustrated by our experiments with salt solutions and glucose solutions. These solutions represent a standard basis of biological samples. To increase sensitivity of our spectroscopical analysis, we normalize the frequency characteristics to the background (when a bias field $\vec{H}_0 = 0$) transmission characteristics.

This work presents first studies of an innovative microwave sensing technique. Many questions arise for further implementation of this technique. One of the questions concerns precise determination of location of a maximal magnetic field (where a ferrite disk is placed) and location of a maximal electric field (where a sample under the test is placed). Since we use an interference method, this problem should, certainly, be considered more in details in further research. Another problem concerns interpretation of actual biological samples. In our experiments, both salt and glucose solutions lead to a single Lorentzian peak, whose position depends on concentration. At the present stage of research, we cannot say definitely what the actual biological samples are. To answer a question on interpretation of actual biological structures and many other important questions, further serious studies with our spectroscopy technique are necessary.

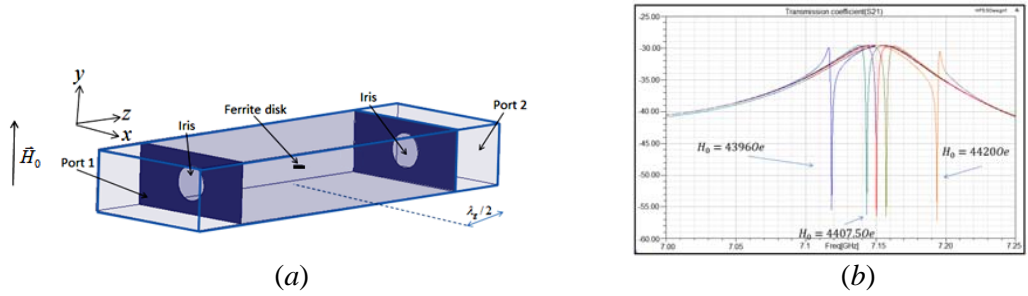

Fig. 1. Interaction the cavity and MDM oscillations. (*a*) A rectangular-waveguide cavity with a thin-film ferrite disk; (*b*) variation of the MDM-resonance lineshapes at variation of a bias magnetic field.

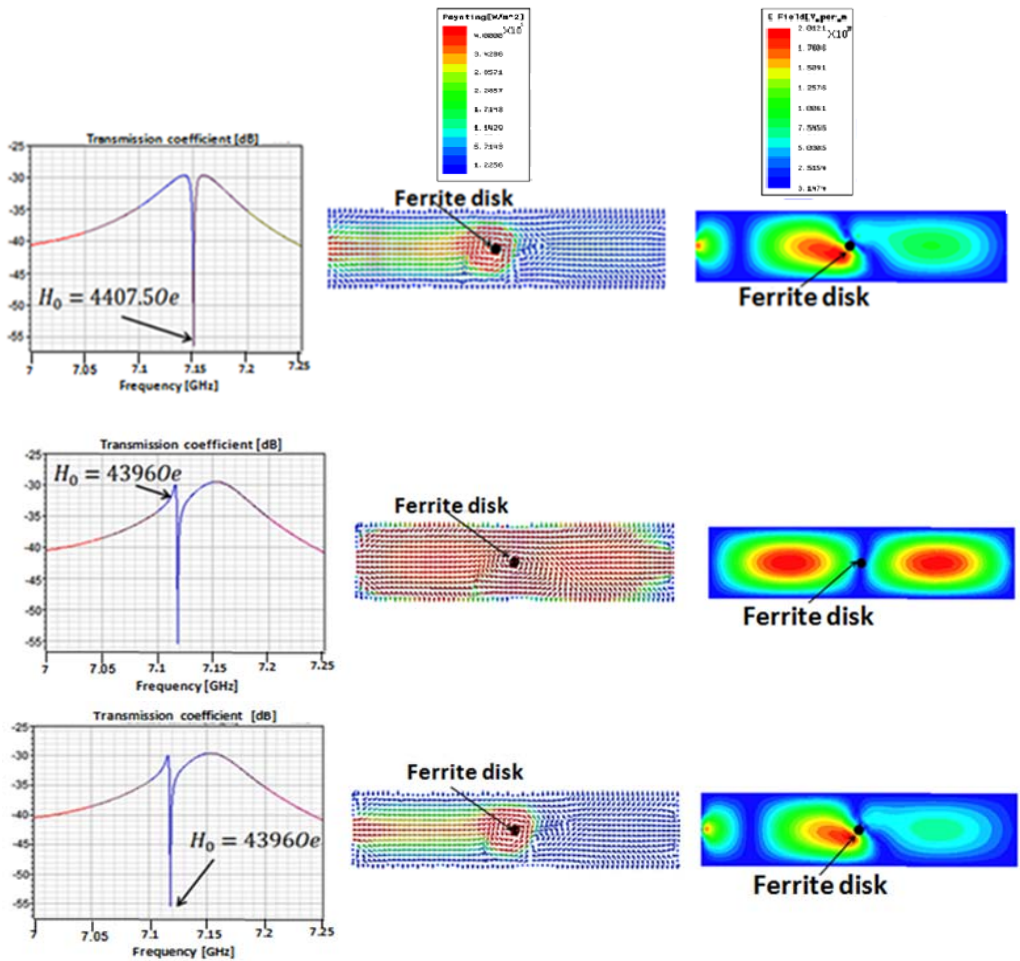

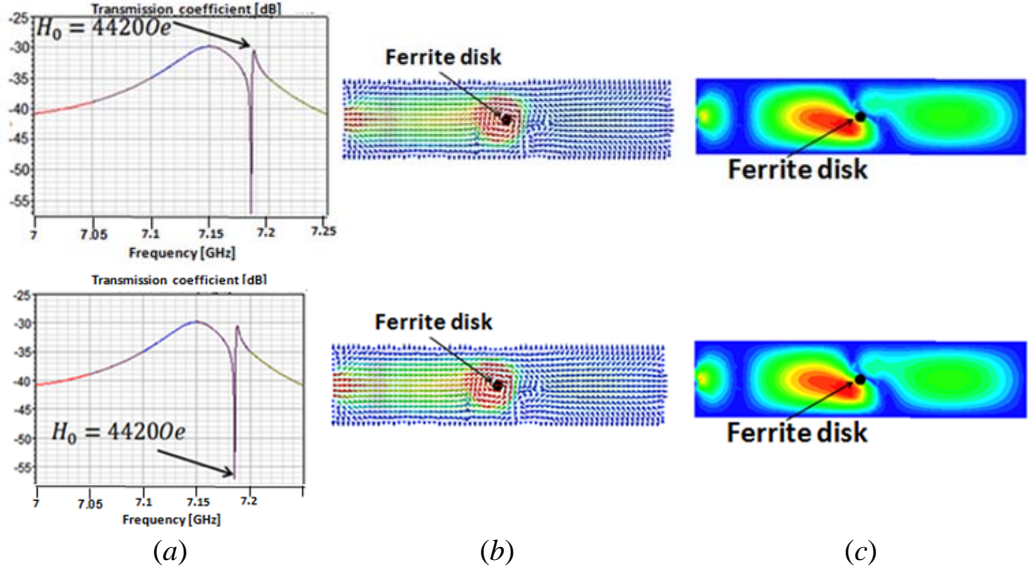

(a)                          (b)                          (c)

Fig. 2. The topological structures of the fields (distributions of the Poynting vector and the electric-field magnitude) in a cavity for different quantities of a bias magnetic field for the Lorentz resonance ($H_0 = 4407.5 Oe$) and Fano resonances ($H_0 = 4396 Oe$ and $4420 Oe$). Column (a): transmission coefficients at different quantities of a bias magnetic field; column (b): the Poynting-vector distributions; column (c): distributions of the electric-field magnitude.

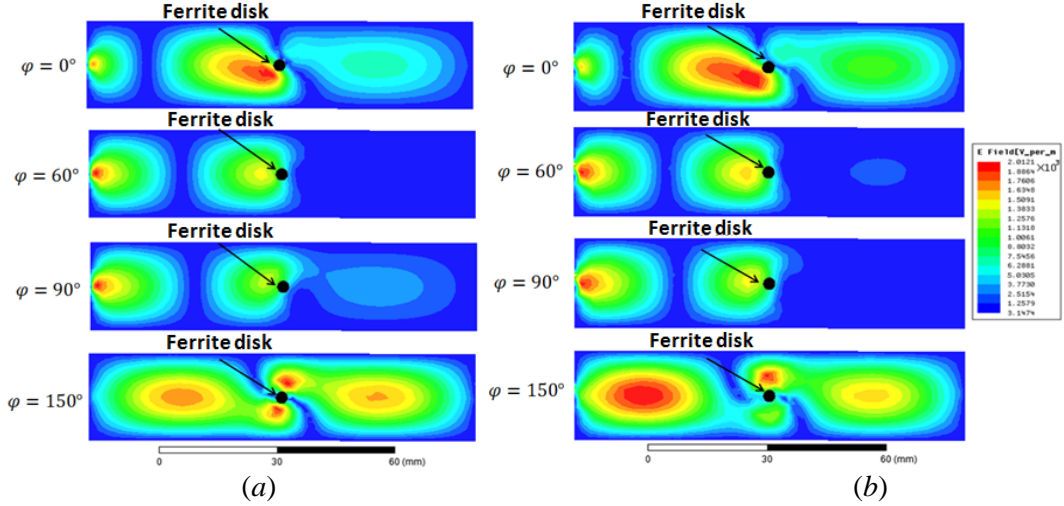

(a)                                    (b)

Fig. 3. Dynamics of the field structure for a case of destructive interference. (a) The electric-field magnitude for different time phases at $H_0 = 4396 Oe$, (b) the electric-field magnitude for different time phases at $H_0 = 4420 Oe$.

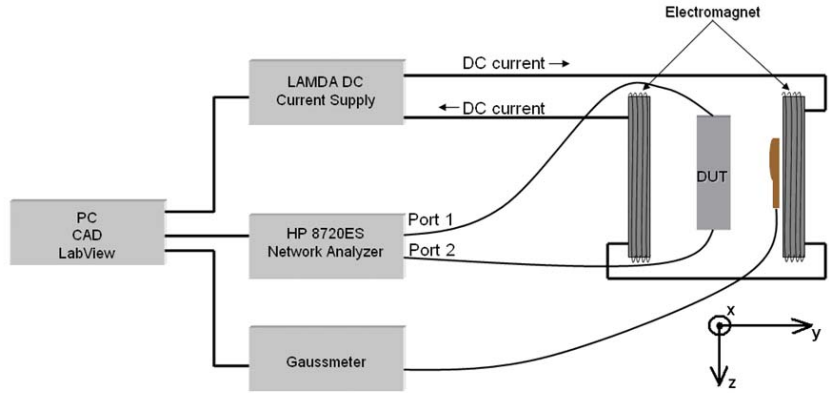

(*a*)

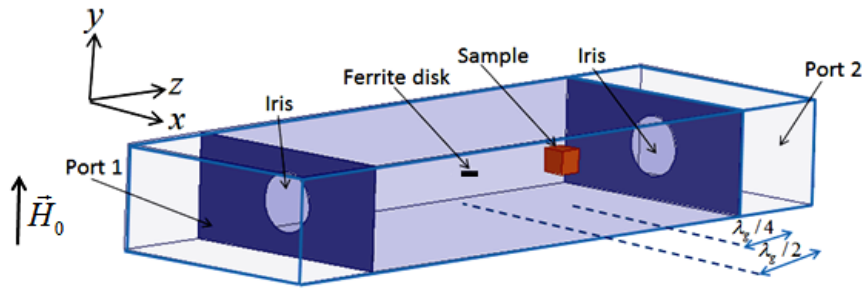

(*b*)

Fig. 4. (*a*) Experimental setup for characterization of high-absorption-matter parameters based on MDM Fano resonances. (*b*) A device-under-test (DUT): a microwave cavity with embedded a small ferrite disk and a material sample.

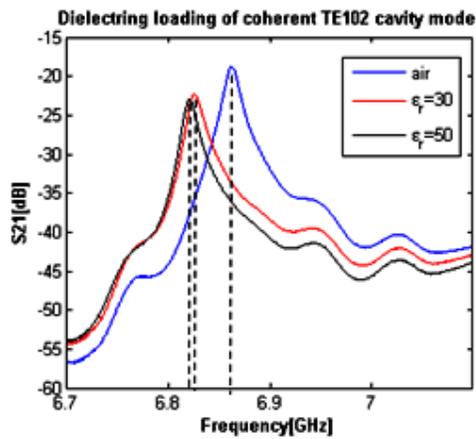 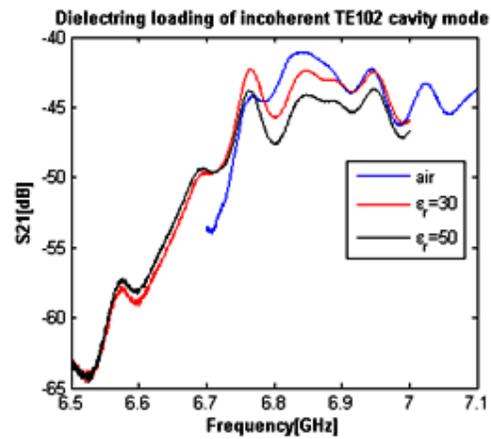

(*a*)                                                                (*b*)

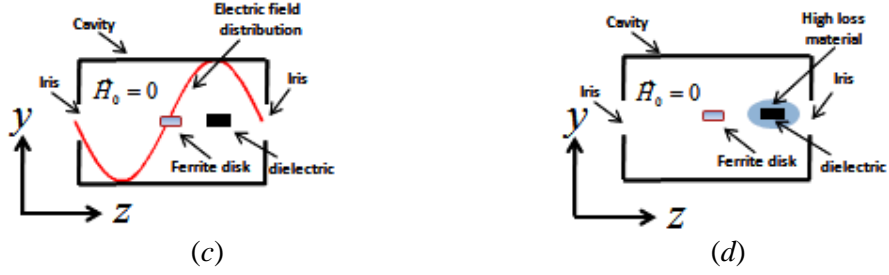

(c)                                                          (d)

Fig. 5. Perturbation of a microwave cavity when a bias magnetic field $\vec{H}_0$ is zero. (*a*) The frequency characteristics of a transmission coefficient when a cavity is loaded with small low-loss dielectric samples; the frequency characteristics of a transmission coefficient when a cavity is loaded with high-loss samples; (*c*) and (*d*) illustration of the experimental structures for low-loss and high-loss samples, respectively.

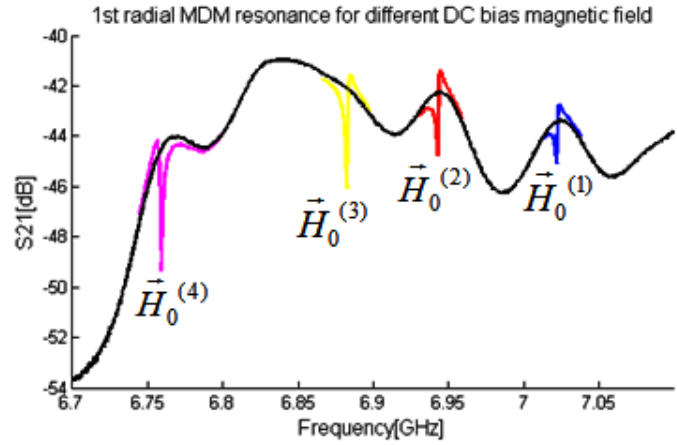

(*a*)

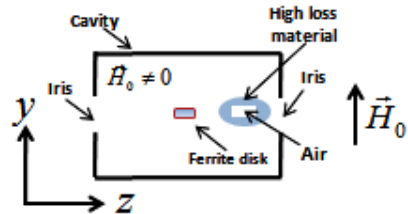

(*b*)

Fig. 6. (*a*) The frequency characteristics of a transmission coefficient when a cavity is loaded with high-lossy wrapping material and a bias magnetic field is applied to a ferrite disk. For magnetic fields $\vec{H}_0^{(1)}$ and $\vec{H}_0^{(2)}$, there the same type of the peak lineshape corresponding to the Fano asymmetry parameter $q > 0$, for a bias field $\vec{H}_0^{(4)}$ there is another type of the peak lineshape with the parameter $q < 0$. A bias magnetic field $\vec{H}_0^{(3)}$ gives us exactly a Lorentzian-like shape ($q = 0$). (*b*) Illustration of the experimental structure.

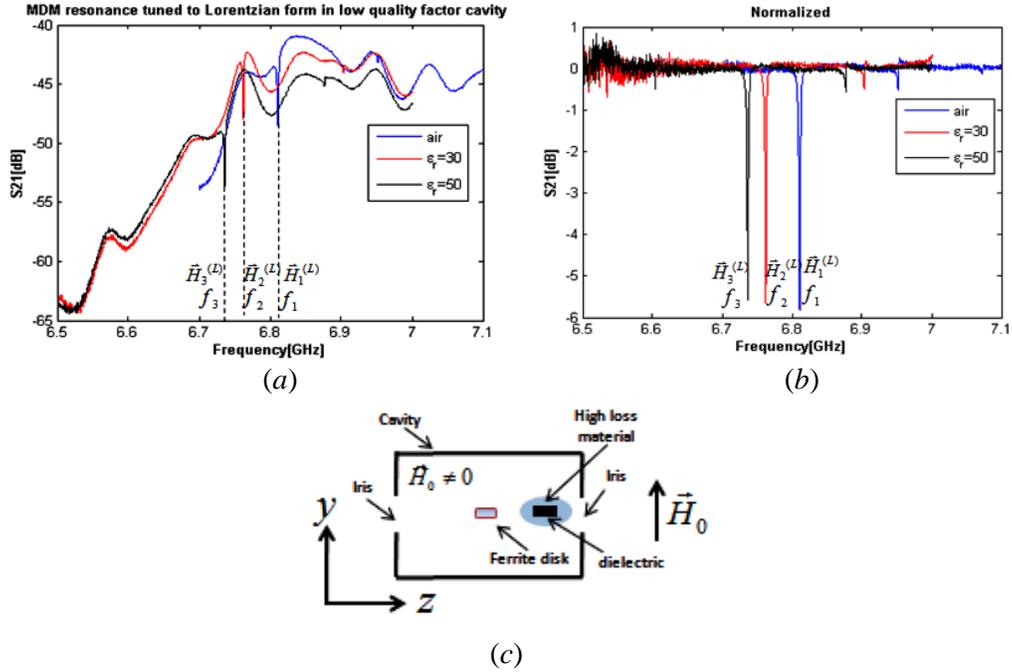

(c)

Fig. 7. Spectroscopical pictures for high-lossy-material samples obtained for the main MDM of a ferrite disk. (a) With variation of a bias field one gets the necessary Lorentz-peak bias field for every type of a high-lossy-material sample. The frequency of such a Lorentz-peak marks exactly the cavity resonance frequency for a certain material loading this cavity. (b) The normalized spectroscopical pictures. (c) Illustration of the experimental structure. The shown frequencies are: $f_1 = 6.811$ GHz, $f_2 = 6.762$ GHz, $f_3 = 6.737$ GHz.

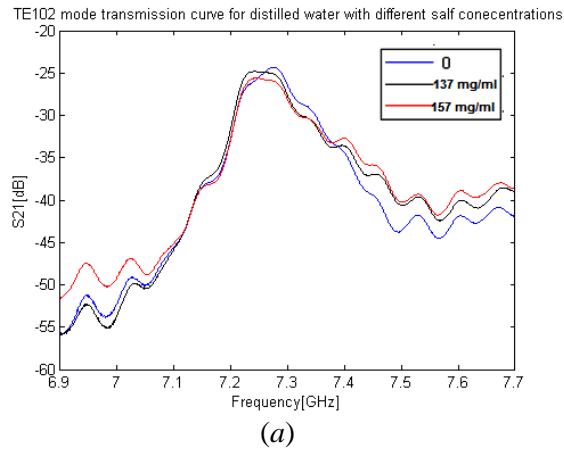

(a)

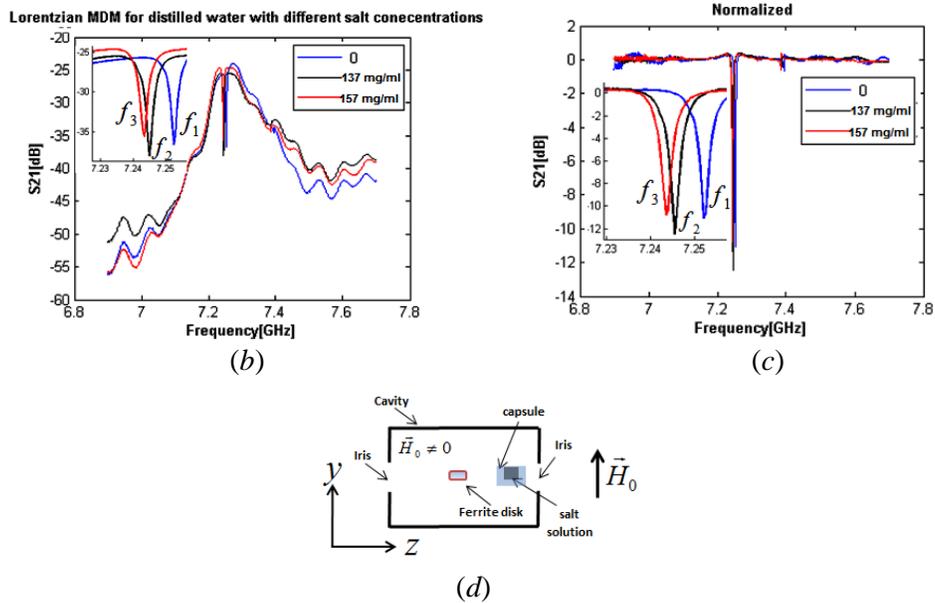

(b)  (c)

(d)

Fig. 8. Fano-interference technique for spectroscopical signature of the salt solitions. (*a*) The frequency characteristics of a transmission coefficient of a cavity loaded by a small capsule with distilled water and salt solutions. The capsule is placed in a maximum of the RF electric field of the cavity and a bias magnetic field is zero. (*b*) The frequency characteristics when a bias magnetic field is applied to a ferrite disk. The Lorentz-peak marks allow exact characterization of the salt concentration. (*c*) The normalized spectroscopical pictures at the same bias magnetic fields. (*d*) Illustration of the experimental structure. The shown frequencies are: $f_1 = 7.252$ GHz, $f_2 = 7.245$ GHz, $f_3 = 7.244$ GHz.

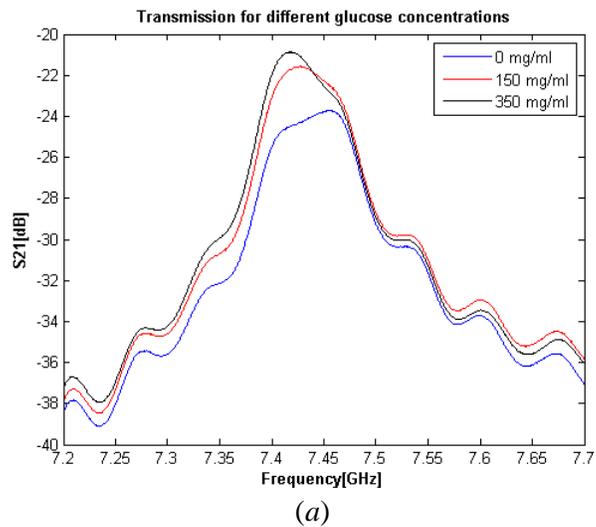

(a)

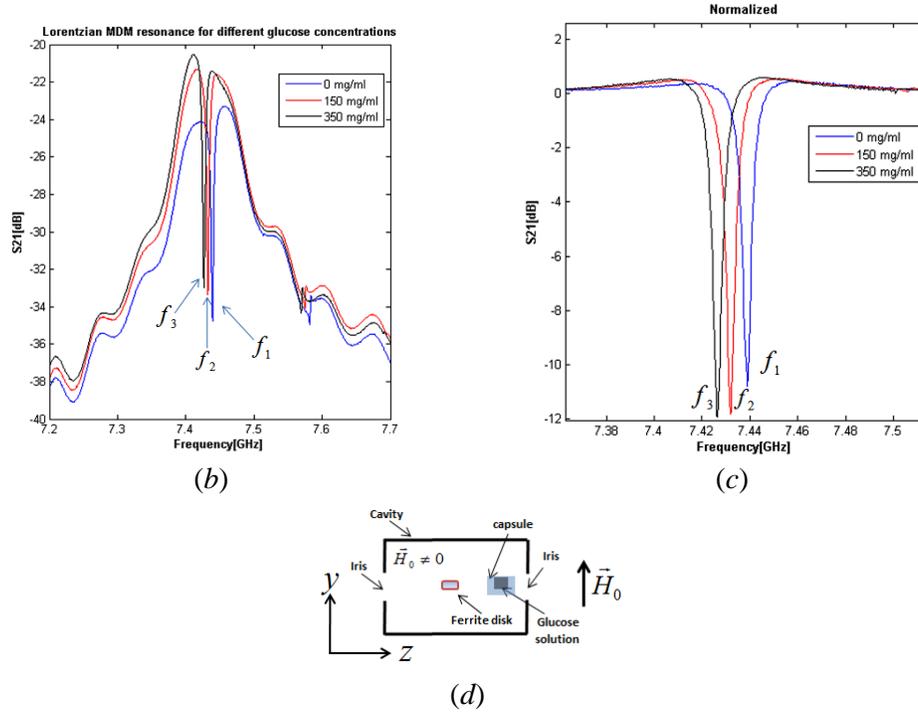

(b)

(c)

(d)

Fig. 9. Fano-interference technique for spectroscopical signature of the glucose solitions. (*a*) The frequency characteristics of a transmission coefficient of a cavity loaded by a small capsule with distilled water and glucose solutions. The capsule is placed in a maximum of the RF electric field of the cavity and a bias magnetic field is zero. (*b*) The frequency characteristics when a bias magnetic field is applied to a ferrite disk. The Lorentz-peak marks allow exact characterization of the glucose concentration. (*c*) The normalized spectroscopical pictures at the same bias magnetic fields. (*d*) Illustration of the experimental structure. The shown frequencies are: $f_1 = 7.439 \text{ GHz}$, $f_2 = 7.432 \text{ GHz}$, $f_3 = 7.426 \text{ GHz}$.